\documentclass[a4paper,12pt,oneside]{report}
\usepackage[english]{babel}      % česky psaná práce
\usepackage[utf8x]{inputenc}  % vstupní znaková sada: ISO Latin 2

\usepackage{amsmath} % balíček pro pokročilou matem. sazbu
\usepackage{amssymb}
\usepackage{url}
\usepackage{hyperref}
\usepackage{pdfpages}
\usepackage{breakcites} 
\usepackage{tkz-graph}

\usetikzlibrary{arrows}
\pgfarrowsdeclare{biggertip}{biggertip}{%
  \setlength{\arrowsize}{1pt}  
  \addtolength{\arrowsize}{.5\pgflinewidth}  
  \pgfarrowsrightextend{0}  
  \pgfarrowsleftextend{-5\arrowsize}  
}{%
  \setlength{\arrowsize}{1pt}  
  \addtolength{\arrowsize}{.5\pgflinewidth}  
  \pgfpathmoveto{\pgfpoint{-5\arrowsize}{4\arrowsize}}  
  \pgfpathlineto{\pgfpointorigin}  
  \pgfpathlineto{\pgfpoint{-5\arrowsize}{-4\arrowsize}}  
  \pgfusepathqstroke  
}  
\tikzset{
  LabelStyle/.style = { rectangle, rounded corners, draw,
                        minimum width = 2em,
                        text = red, font = \bfseries },
  VertexStyle/.append style = { inner sep=5pt,
                                font = \Large\bfseries},
  EdgeStyle/.append style = {->, arrowhead=1em} }

\newcommand{\ignore}[1]{}
\newcommand*{\defeq}{\stackrel{\text{def}}{=}}

\newcommand{\cvut}{Czech Technical University in~Prague}
\newcommand{\fjfi}{Faculty of Nuclear Sciences and Physical Engineering}

\newcommand{\autor}{Ing. Martin Plajner}           % zde VYPLŇTE své jméno a příjmení
\newcommand{\rok}{2016}                % zde VYPLŇTE rok odevzdání, např. 2006
\newcommand{\vedouci}{Ing. Jiří Vomlel, Ph.D.}         % zde VYPLŇTE jméno a příjmení vedoucího práce, včetně titulů
\newcommand{\keyword}{computerized, adaptive, testing, Bayesian networks, IRT, CAT, neural networks}       % zde NAPIŠTE anglicky max. 5 klíčových slov (přeložte z~češtiny)
\newcommand{\abstrEN}{In this paper we follow our previous research in the area of Computerized Adaptive Testing (CAT). We present three different methods for CAT. One of them, the item response theory, is a well established method, while the other two, Bayesian and neural networks, are new in the area of educational testing. In the first part of this paper, we present the concept of CAT and its advantages and disadvantages. We collected data from paper tests performed with grammar school students. We provide the summary of data used for our experiments in the second part. Next, we present three different model types for CAT. They are based on the item response theory, Bayesian networks, and neural networks. The general theory associated with each type is briefly explained and the utilization of these models for CAT is analyzed. Future research is outlined in the concluding part of the paper. It shows many interesting research paths that are important not only for CAT but also for other areas of artificial intelligence.}                  % zde NAPIŠTE abstrakt v~angličtině

\begin{document}

\thispagestyle{empty}

\begin{center}
    \Large  \textbf{\cvut}\\[2mm] \fjfi 
		
    \vspace{10mm}

    % logo CVUT -- pokud jej nechcete použít, zakomentujte následující řádek a odkomentujte řádek pod ním:
      \vspace{1mm}
    
		\begin{center}
		{\includegraphics[height=25mm]{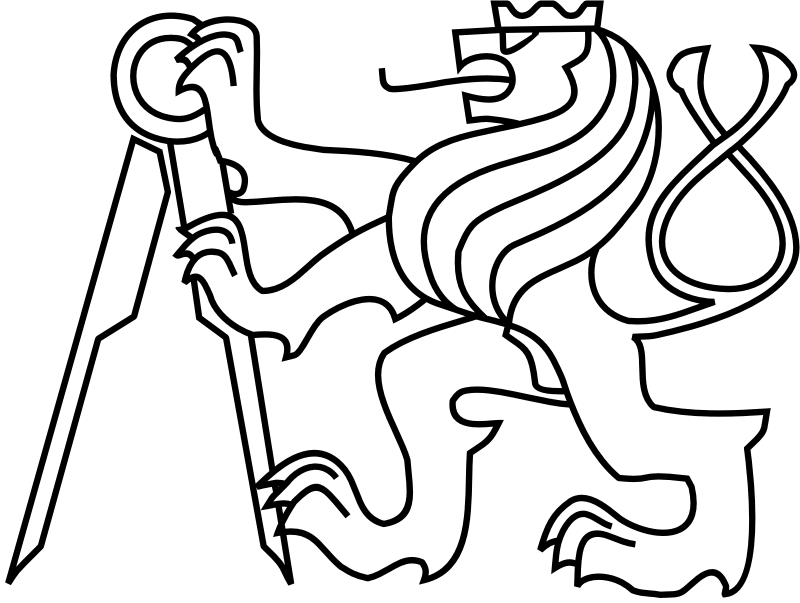}}
		\end{center}
    \vspace{7mm}
   %\vspace{50mm}

   \huge \textbf{Probabilistic Models for Computerized Adaptive Testing}
	
   \vspace{5mm}
   \Large Study for dissertation thesis
	\vspace{5mm}
	\normalsize
	
  \textbf{Abstract:}  	
	\end{center}
	\abstrEN  
	\begin{center}
	\vspace{1em}
   
	\vfill
	
   {\large
    \begin{tabular}{rl}
    Author: & \autor\\
    Supervisor: & \vedouci\\
    Year: & \rok
    \end{tabular}
   }
\end{center}

\newpage   
\tableofcontents 

\newpage % SEM NESAHEJTE!

\chapter*{Introduction} \addcontentsline{toc}{chapter}{Introduction} 
Educational testing is an important part of our lives in the modern society. Every person participates in a large number of tests which are used to assess his/her level of knowledge, quality, or skill in a certain domain. There are many different possibilities how to design a test for a specific purpose. The theory of test creation, administration, validation, etc. (in general called psychometrics) is extensive (for example,~\cite{1964psychometrics, Lord1968, Rasch1960, Mislevy1994}, and many others). The process of the creation of a good test is long, contains many steps, and can be performed in many different ways. In the classical approach we first identify the target ability (the ability we want to measure). Afterwards, there are repeated cycles of adding new questions, testing the test on a small set of examinees (to see if the questions are measuring the right ability in a correct way), and removing unsuitable questions. In the end we end up with one satisfactory version of the test. This test is fixed in its questions (i.e., every student taking the test will have the same question). This approach does not take into account the individuality of each examinee. It is clear that for a skilled examinee the test will necessarily contain a lot of questions which are too easy and vice versa. The time which is being spent by solving these questions could be used to better explore his/her actual skill level. This can be achieved by asking questions with an appropriate difficulty.
 
There are banks of questions which are suitable to measure the ability of a student (sometimes they are quite limited – for example if we measure a physical ability there might be a limited number of possible questions – and sometimes they are unlimited – in mathematics we can create as many different problems to solve as we want). Questions for a test are selected from this bank. If we use one set of questions for every test there will be a lot of possibly good questions which are never asked (those which remained in the question bank unselected). The same set of questions for every test also, in some cases, encourages cheating (which of course can usually be solved by other methods, but it requires additional steps). 

Some tests do not follow the outline explained in the paragraph above and tries to utilize the whole bank of questions. One way of doing that is for example used in the Czech driving license test\footnote{\url{http://www.mdcr.cz/cs/Silnicni_doprava/etesty/etesty.htm}(Czech language)}. It is a computer test where 25 questions are randomly selected from approximately a thousand of possible questions. This approach negates the possibility of learning all the questions by heart as well as cheating by looking into your neighbor’s sheet. Another test in a similar manner is done by the Faculty of Medicine of the Charles University\footnote{\url{http://www.lf1.cuni.cz/prijimaci-rizeni}(Czech language)} as an entrance exam test. Questions for the entrance exam are selected from a set (book) of possible questions for each test (i.e., question bank). Several versions of a test are prepared for every entrance exam session. Both of these selection processes (driver’s license and entrance exam) remove some complications mentioned above but produce new problems. In the driver’s license test, where the question selection is done automatically by the system, it is hard to ensure the overall difficulty of each test will be approximately the same. Cases where a lot of easy questions, or on the contrary a lot of difficult questions, is selected might occur. In the approach of the medical faculty this unfair combination can be avoided by a careful test composition. The test composition is done by hand by specialists. These specialists sometimes have shifted notion of the difficulty of individual questions (some things they may think of as very easy are actually hard for young students). There is also definitely a lot of effort and time involved in the preparation of every entrance exam round.

Computerized Adaptive Testing (CAT) offers a way to overcome some of the limitations given by the classical testing approach. The examinee is answering questions presented to him/her by a computer system. This system is centered on a student model. There are many ways to construct a student model. One way is a model composition by experts. Another is to construct the model from a data set of many previously tested examinees. These examinees have to be tested without the adaptive approach to obtain a basis for the model creation. Afterwards, the model can be further updated and extended with new cases even while being in use. 

During the course of testing the student model is updated to reflect abilities of the tested student and as a part of that process an estimate of student’s level of knowledge is updated as well. This provides us an actual estimate of student’s abilities in every phase of testing. At the same time the model is used to select the next question. The next selected question is the most appropriate one. An appropriate question suits certain criteria, usually providing the best information about the student at the current stage of testing. Questions are selected from a bank of questions. This bank can be similar to a question bank for the classical test. Adaptive testing is performed until a criterion is reached. There is a variety of possible criteria. Usually we want to stop the test when the confidence of the estimate of student’s skill is above a certain significant value. Other practical limitations might affect this criterion such as the total time of the test or the number of asked questions. The adaptive testing concept brings many advantages but also some disadvantages over the classical testing approach. These aspects are detailed in the following chapters. Further we present three different model types for CAT. Experimental results of these models with empirical data are in an associated paper~\cite{Plajner2016}

\chapter{Computerized Adaptive Testing}
This chapter introduces the concept of Computerized Adaptive Testing (CAT) and summarizes its advantages and disadvantages.
 
CAT is a concept of testing which is getting large scientific attention for about two decades~\cite{Linden2000, Wainer2015, VanderLinden2010}. With CAT we build computer administered and computer controlled tests. The computer system is selecting questions for a student taking the test and evaluating his/her performance.

The process can be divided into two phases: model creation and testing. In the first one the student model is created while in the second one the model is used to actually test examinees. There are many different model types~\cite{Almond1999, Culbertson2014, cowell1999} which can be used for adaptive testing. In this work we are going to cover Item Response Theory (IRT), which is a model regularly used for CAT, Bayesian and neural networks (BNs and NNs), which are both models commonly used in many areas of artificial intelligence for a large variety of tasks. We will pay closer attention to these models later on but regardless of the model we choose the testing part follows always the same scheme. With the prepared and calibrated model, CAT testing repeats following steps.

\label{sec:CATprocess}
\begin{itemize}
	\item The next question to be asked is selected.
	\item This question is asked and an answer is obtained.
	\item This answer is inserted into the model.
	\item The model (which provides estimates of the student's skills) is updated.
	\item (optional) Answers to all questions are estimated given the current estimates of student's skills.
\end{itemize}

This procedure is repeated until we reach a termination criterion. There are many different stopping criteria. It can be a time restriction, the number of questions, or a confidence interval of the estimated variables (i.e., reliability of the test).

\section{Advantages of CAT}
\emph{Shorter tests:} One of the most obvious advantages of CAT is that the overall length of a test is reduced. Because questions are selected according to the level of the tested student he/she is not forced to answer questions which are too easy or too hard. This means the test aims better at discovering the level of the student. That results in the reduction of the length of the test in both time and the number of questions. Usually it is enough to ask as few as half the questions to obtain reliable results. 

\emph{Fairness:} A test in the classical theory usually expects a Gaussian score distribution among the population of students. This expectation yields frequencies of question difficulties to be of the same distribution (most questions are medium difficulty and less of them are hard or easy). Because of that a precision of the resulting score is the best for mediocre students while it drops for students on edges of the scale. CAT on the other hand selects appropriate questions based on the skill of the student. That results in the same precision for each student, nevertheless his/her position on the score scale. This topic is further discussed in~\cite{Pine1978}.

\emph{Intelligent tutoring system:} It is quite easy to convert a CAT test to an intelligent tutoring system. ITS is a system which is designed to uncover student’s weak and strong spots and offer more exercises and materials to learn from.

\emph{Motivation: }While testing a student with a CAT system the optimal probability of successful answer to a question is 50\% (at least while using the IRT student model). Even though a question with such probability may not exists to be selected in every step of the testing it should not get far from this value if the question bank is well designed. This helps to keep a student interested in the test. A weaker students will not get overwhelmed by many difficult questions while a good student will not get bored by easy ones.

\emph{Reseating the exam: }With CAT it is extremely easy to resit the exam (provided we keep track of previous questions for the particular student). Because of its nature CAT system can create a completely different test to retest the same student.

\emph{Computer administration:} The test is done electronically and thus results are available immediately and can be stored easily. It is also possible to deliver the test over the internet.

\section{Disadvantages of CAT}
\emph{Over usage of some items: }This issue greatly depends on the way we use to select subsequent questions for students. Nevertheless, with most commonly used criteria there is a danger of selecting the same questions for groups of students and/or selecting certain questions in many tests. For example, the first question, if the selection process is not modified, will be the same for each student. We have no information about the student so far and the selection process results in the same question. Following questions will be the same for groups of students. These groups shrink with more answered questions as the number of possible combinations of answers increase. This behavior can be reduced by having a large question bank containing many different questions with similar properties (i.e., difficulty). Moreover, it is possible to modify the selection process to ensure a wider spread of selected questions over the question bank (with the cost of decreased precision).

\emph{Initial data collection:} Prior to starting a test using CAT it is necessary to obtain a large set of data (full test results) from a representative population. This data is used to create and calibrate the student model used for testing. Results used for this creation need to come from a full length tests (optionally it would be possible, but not preferable, to have a several sub-tests). This means students participating in the initial testing are required to fill answers to many items. Another option is to build a model with the help of an expert in the field but even this approach is time consuming.

\emph{Computer administration:} In order to test students it is necessary to create an environment for such testing on the computer. Also it is necessary for students to have access to a computer rather than having just a pen.

\emph{Results perception:} Last but not least, there might be some issues with the perception of results by students taking the test. It may be hard to explain to them and for them to comprehend the fact, that even though they got completely (or partly) different questions they are sorted on the same scale (sometimes even obtaining the same score). It may seem unfair and incomparable because of the question selection process. The feeling may be the same as with the the Czech driving license test mention in the introduction but there the selection is done at random. In reality CAT tests tend to be more fair the regular paper-pen tests~\cite{Moe1988, Tonidandel2002}.

\chapter{Data Collection}
 
To support the creation of a student model we have collected empirical data. We designed a paper test of mathematical knowledge of grammar school students. The test focuses on simple functions (mostly polynomial, trigonometric, and exponential/logarithmic). Students were asked to solve various mathematical problems\footnote{In this case we use the term mathematical ``problem'' due to its nature. In general tests,	 terms ``question'' or ``item'' are often used. In this article all of these terms are interchangeable.} including graph drawing and reading, calculating points on the graph, root finding, describing function shapes and other function properties. 

\section{Test Design}
When we were creating the test for the data collection we performed several steps. First, we prepared an initial version of the test. This version was carried out by a small group of students and took about 80 minutes to be solved. We evaluated this first version and based on this evaluation we made changes before the main test cycle. It was necessary to limit the time of the test to 45 minutes to make it fit one school lesson. Some questions were removed completely from the test. They were mainly those where the information benefit of the problem was too low due to their high or low difficulty (i.e. only a few students answered them correctly or incorrectly). There was no assumption that all the students should be able to finish all the questions in time which is a usual way to create school test. In this case we were targeting the number of questions to allow the best students to finish just in time. This allowed us to remove less questions than we normally would. Remaining problems were updated and changed to be better understandable.  Moreover we divided problems into subproblems in the way that: 
\begin{itemize}
\item[(a)] it is possible to separate the subproblem from the main problem and solve it independently or 
\item[(b)] it is not possible to separate the subproblem, but it represents a subroutine of the main problem solution. 
\end{itemize}
Note that each subproblem of the first type can be viewed as a completely separate problem. 
On the other hand, subproblems of the second type are inseparable pieces of a problem.\\

The final version of the test contains 29 mathematical problems. These problems have been further divided into 53 subproblems. Subproblems are graded so that the sum of their grades is the grade of the parent problem, i.e., it falls into the set $\{0,\ldots,4\}$. Usually a question is divided into two parts each graded by at most two points\footnote{There is one exception from this rule: The first problem is very simple and it is divided into 8 parts, each graded by zero or one point (summing to the total maximum of 8 points).}. The granularity of subproblems is not the same for all of them and is a subset of the set $\{0,\ldots,4\}$. All together, the maximal possible score in the test is 120 points. \\
In an alternative evaluation approach, each subproblem is evaluated using the Boolean values (correct/wrong). An answer is evaluated as correct only if the solution of the subproblem and the solution method is correct unless there is an obvious numerical mistake. 

We organized tests at four grammar schools. In total 281 students participated in testing. In addition to answers to questions, information about students was collected. This includes mostly some personal factors as gender, age, and grades from mathematics, physics, and chemistry from the recent period. These factors will be used to better differentiate between students and to better predict their performance as well as to verify the validity of the test. The goal of the test was to pinpoint students' weak and strong points and to provide them with valuable information about their skills. Students are able to view their results (the scores obtained in each individual question). We also provide them with a comparison with specific groups of students (their class, school, and all participants). Comparisons are provided in the form of quantiles in the respective group.

\section{Test Assessment}
In the following section we present a psychometric analysis of the test. This kind of analysis should be done for every large scale test. It might not be necessary to perform all actions which are presented below for CAT. Nevertheless we will use these results to compare classical approach and CAT as well as to point out some interesting relations. Moreover it proves that the paper test we used to collect data provide reasonable results.

\textbf{True scores and reliability}\\
The goal of every test is to measure a certain variable. This variable reflects examinee's skill, ability or level of another quality (some psychiatric test might be measuring person's empathy). In terms of IRT and CAT this variable is a part of the student model described in the Section~\ref{sec_IRT}. Even in the classical test there is a certain variable. A test is just a tool created to measure this variable. As always, when measuring anything, the measurement process is obstructed with measurement errors. These errors are caused by many different factors (the examinee could have a bad day, be ill, guess the answer, or get distracted while solving a single question,\ldots) and it is reasonable to expect them to have a significant influence on the final value. The value obtained as a measurement $x$ of the variable $X$ is called a raw score and is in the form
$$x = \tau + e$$
where $\tau$ is the true score and $e$ is an additive error. 

There is an obvious question whether the raw score is influenced more by the true score or the error. For many measurements the maximum-likelihood estimator of the error is the variance of many consecutive measurements of the same factor. In our case it proves to be impractical to measure one person multiple times for obvious reasons. It is not as well possible to use the variance of many different examinees as their true values most likely differ. The variability of scores in the data set is then caused by actual differences between examinees (different true scores) as well as errors. It is usually expected that the data set satisfies homoskedasticity condition\footnote{Homoskedasticity means that the size of an error is not correlated with the size of the measured variable}. With this assumption true scores and errors are statistically independent and thus the observed variance $\sigma_x$ is a sum of variances of true scores $\sigma_\tau$ and errors $\sigma_e$.
$$\sigma_x=\sigma_\tau+\sigma_e$$
The best possible situation is that the variance of the measured variable X is fully modeled by true scores. This situation is very unlikely to happen. To determine the level of the relationship we use the value called reliability\footnote{Note that reliability is a well established and very important property of a test among psychometric comunity} which is defined as follows:
$$r_{xx} = \frac{\sigma_\tau}{\sigma_x}=\frac{\sigma_\tau}{\sigma_\tau+\sigma_e}$$
The higher the value the better. Unfortunately variables $\sigma_\tau$ in the nominator as well as $\sigma_e$ in the denominator of the second fraction are hidden (unobservable) variables and as such we are unable to evaluate their variance. The reliability has to be estimated with a different approach. 

There are many possible approaches and we will elaborate more into one of them which is known as Cronbach’s alpha coefficient. The idea is that items of the test are measuring the same factor and thus they should correlate with each other. The amount of pair wise correlations for $q$ questions is ${k=\frac{q(q-1)}{2}}$. All these correlations are put together in the Cronbach’s alpha coefficient which can be calculated as
$$r_{xx}\approx\alpha=\frac{n}{n-1}\left(1-\frac{\sum_{i=1}^{n}\sigma_i^2}{\sigma_t^2}\right)$$
where $\sigma_i$ is the variance of the ith item of the test, $\sigma_t$ is the variance of the whole test and $n$ is the number of items in the test. The coefficient should reach high values. According to \cite{1964psychometrics} any value below 0.5 means the test is of no use. Quality results are produced with the coefficient over 0.9.

For our data set (281 students) the following values were calculated:\\
Cronbach’s alpha for the numeric classification: $\alpha = 0.914$\\
Cronbach’s alpha for the Boolean classification: $\alpha = 0.925$\\
These values show reasonably high reliability of the test.

\textbf{Normalization and standard scores}\\
It may not be very efficient to use directly the score a student obtained in the test. This score is called the raw score. The question with raw score is that it may not distinguish between individual students as well as it could. For example in case we have 3 results with scores of 20, 40 and 60 respectively. It would seem to us that the gap between the first and the second pair is the same. It definitely is in terms of raw score, but it may not be in terms of real abilities of students. If there are a lot of students who score between 20 and 40 and just a few in the interval between 40 and 60 then the skill gap from the second to the third may not be as wide as it appears. In order to better categorize students, scores are usually normalized. With normalized scores it is easier to evaluate the position of a student in the test for a specialist who is used to work with normalized scores. There are many different types of standard scores and most of them are obtained by a linear transformation of raw scores (note that it means that the order of examinees is not changed by this kind of transformation) by the following formula
$$x'=\mu'+\sigma'\frac{(x-\mu)}{\sigma}$$  
Where $x'$ is the transformed score, $\mu'$ and $\sigma'$ are desired mean and variance values of the standardized score, $\mu$ and $\sigma$ are previous mean and variance values and $x$ is the raw score. 

To apply these transformations it is required that the raw score belong to the Gaussian distribution (ideally with the mean value in the middle of possible scores). Standardized scores differ in the chosen parameters of $\mu'$ and $\sigma'$ and some special selections are generally recognized. The most commonly used is the z-score with the mean value~0 and the variance~1. Another well known standard score is the IQ~score (${\mu' = 100}$, ${\sigma'=15}$) used mostly for intelligence testing. Other well known scores are also stens, stenines, percentiles, and t-scores.

\begin{figure}%
\begin{center}
\includegraphics[width=0.8\columnwidth]{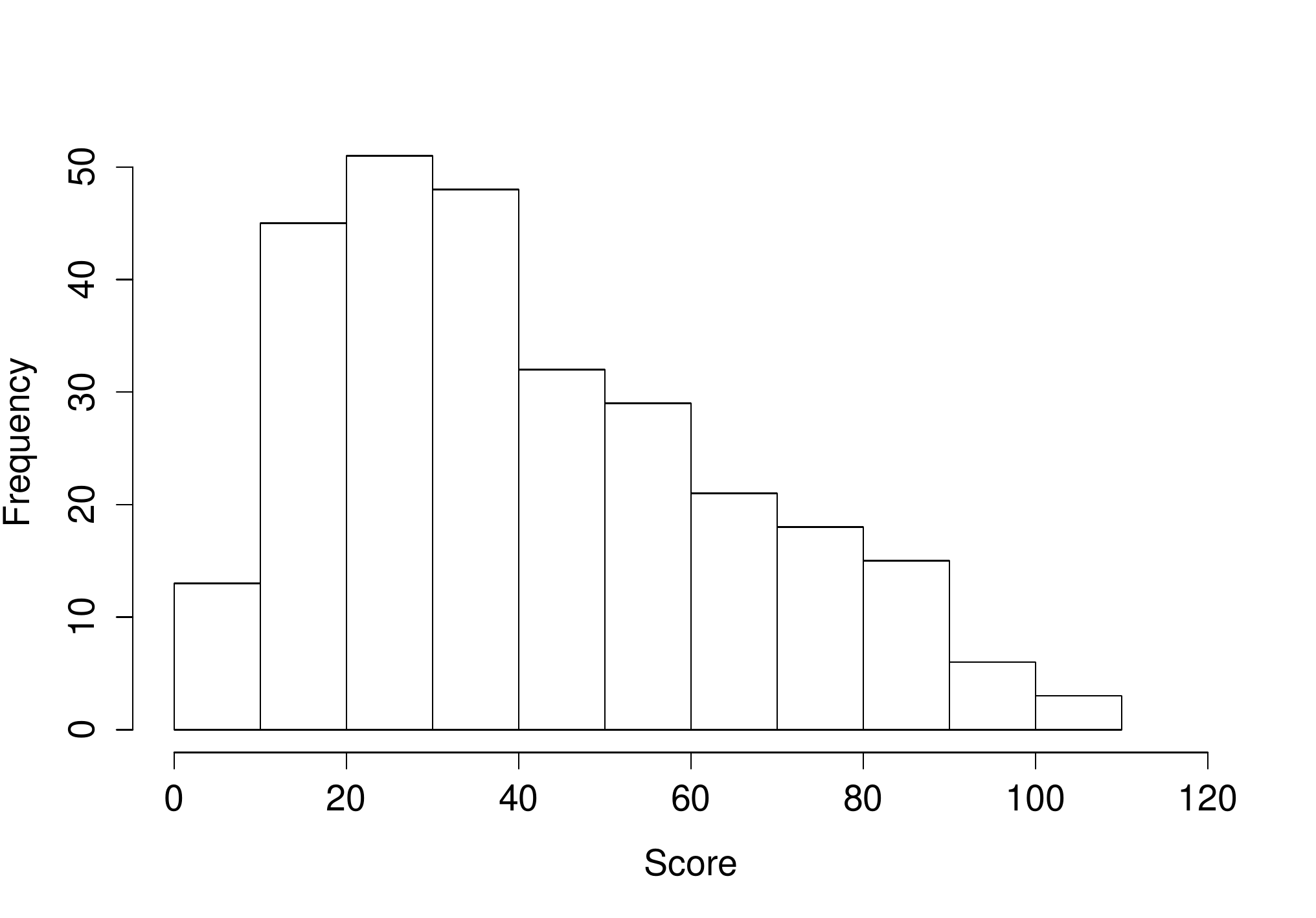}%
\caption{Score frequencies}%
\label{pic:gauss}%
\end{center}
\end{figure}

The set of scores obtained from our data set most likely do not belong to Gaussian distribution. The visual proof is displayed in the Figure~\ref{pic:gauss} where it can be clearly seen that it does not resemble the Gaussian distribution. The Shapiro-Wilk normality test also rejects the null hypothesis of the Gaussian distribution by resulting with $p-value = 3.648\cdot^{-7}$. The solution to this problem is provided by the McCall’s area standardization~\cite{McCall1922, 2011psychometrics} which transforms raw scores to the Gaussian distribution. We performed this step at first and then we transformed scores to the standardized score scales. To illustrate these scales, a short excerpt from whole scale tables for the z-score and the IQ score is shown in the Table~\ref{tab:scores}. From this table we can see that the center of the normalized score scale is around 40 points of raw score (z value of 0 and IQ of 100). Maximum score obtained was 107 points and minimum 0. The transformed scale at these points has opposite (and extreme) values. The space between center of 40 points and these extremes is the same at both sides for normalized scores but it is not for raw scores. There is the same amount of students scoring in any two intervals of the same length ending/starting at the center on the normalized score scale (for example the same amount of students scored in intervals (-1,0) and (0,1) on the z-score, which corresponds to raw scores of (20, 40) and (40, 72 - not in the table)).

% Table generated by Excel2LaTeX from sheet 'List1'
\begin{table}[htbp]
  \centering
  \caption{Standardized scores}
	\label{tab:scores}
    \begin{tabular}{lrrrrrrr}
		\hline
    raw&0&20&40&60&80&100&107\\
    \hline
    z&-2.91&-1.00&-0.01&0.67&1.16&2.06&2.91\\
    IQ&56&85&100&110&117&131&144\\
    \hline
    \end{tabular}%
  \label{tab:addlabel}%
\end{table}%

\textbf{Validity}\\
Another question it is important to ask is whether the test is actually measuring the factor it is supposed to measure (i.e. in our case if the score obtained reflects mathematical skills rather than for example the ability to read the question or the writing skill of the examinee). This characteristic is called validity and there are many different ways of proving the test is valid. Most validity proofs come from the outside of the test. One way is to let an examinee to answer a new different test measuring the same factor (ideally a test which is already well established). Another way is to consult other factors known about the examinee, which is what was performed in our case.

As was mentioned above, in addition to solutions to individual problems student's grades from subjects (mathematics, physics, and chemistry) were obtained. It is reasonable to expect a correlation between these grades and the score reached. The correlation is present and its values are shown in the following paragraphs. Because of this fact, although the complete validation would require more thorough examination, it is expected that the test is valid.

\section{Preliminary Test Statistics}
In this section we present an overview of results obtained in testing. This should provide an idea of skills of students, prove validity as mentioned in the paragraph above and will also be referenced later on for comparison.

\begin{table}[htbp]
  \centering
  \caption{Average test scores of the four grammar schools.}
	
	\vspace{3mm}
	
    \begin{tabular}{l|rrrr|r|}
		\hline
    &\textbf{GS1}   & \textbf{GS2} & \textbf{GS3} & \textbf{GS4} & \textbf{Total} \\
					\hline    
    \textbf{Males} & 51.40  & 40.08 & 47.77 & 51.03 & 48.48 \\
    \textbf{Females} & 42.53 & 54.86 & 44.45 & 38.81 & 43.06 \\
		\textbf{Together}&42.76 & 46.68 & 46.35 & 43.65 & 44.53 \\
		\hline
    \end{tabular}%
  \label{tab:totals}%
\end{table}%

The Table~\ref{tab:totals} shows the reached scores divided by gender and school. We calculated Pearson's correlation coefficients of score with other factors. Results are shown in the Table~\ref{tab:corr1}. The correlation test is associated with its p-value, where the null hypothesis is correlation of 0 (no correlation). It means we can say that the correlation between score and all grades (math, physics, and chemistry) is present. The negative value of correlation means that better grade (lower value) yields better score (higher value) which is expected. Furthermore we can see that the grade in mathematics has the highest correlation while physics and chemistry lower. Another significant correlation is interestingly between the fact that the student filled his/her name and his/her score. Positive value shows that those students who filled their name scored better in the test.
On the other hand we can not reject the null hypothesis for gender, so there most likely is no statistically significant correlation between gender and score\footnote{Females were encoded as 1 and males as -1. Negative value would show worst score for females, but it is statistically insignificant.}.

\begin{table}[htbp]%
\caption{Correlations of the score with other factors}
\label{tab:corr1}
\begin{center}
    \begin{tabular}{rrrrrr}
    \hline
    \multicolumn{1}{l}{\textbf{}}  & \multicolumn{1}{l}{\textbf{Gender}} & \multicolumn{1}{l}{\textbf{Mathematics}} & \multicolumn{1}{l}{\textbf{Physics}} & \multicolumn{1}{l}{\textbf{Chemistry}}& \multicolumn{1}{l}{\textbf{Name}} \\
    \hline
    \textbf{Correlation}&  -0.10  & -0.59  & -0.42  & -0.41  & 0.22\\
		\textbf{p-value}&  0.08  & 2.20E-16  & 3.63E-12 & 2.65E-11 & 0.18E-4 \\
    \hline
    \end{tabular}%
\end{center}
\end{table}

Some questions were in the form of word problems with a connection to everyday life (calculating savings, time to finish a job,etc.). These questions were correlated with the score independently as well. The result is displayed in the Table~\ref{tab:corr2}. In the first column it is possible to see that there is a strong and statistically significant correlation of the score obtained in these questions with the total score. Also in this case there is not a strong correlation with the gender of the student even though a bit higher and on the edge of rejection of statistical insignificance. The trend of correlations with grades is preserved but the strength of correlation is lower. In connection with previous results, it leads to an assumption that students with worse grades from these subjects answered correctly rather this kind of questions than other questions.

\begin{table}[htbp]%
\caption{Correlations of word problems with other factors}
\label{tab:corr2}
\begin{center}
    \begin{tabular}{rrrrrr}
    \hline
    \multicolumn{1}{l}{\textbf{}} & \multicolumn{1}{l}{\textbf{Score}} & \multicolumn{1}{l}{\textbf{Gender}} & \multicolumn{1}{l}{\textbf{Mathematics}} & \multicolumn{1}{l}{\textbf{Physics}} & \multicolumn{1}{l}{\textbf{Chemistry}} \\
    \hline
    \textbf{Correlation}& 0.69  & -0.19  & -0.38  & -0.25  & -0.27  \\
		\textbf{p-value}& 2.20E-16   & 0.16E-3  & 3.16E-10  & 7.99E-5   & 2.25E-5   \\
    \hline
    \end{tabular}%
\end{center}
\end{table}

\chapter{Models for Adaptive Testing}

We remind, as was mentioned in the Section~\ref{sec:CATprocess}, the process of an adaptive test.
\begin{itemize}
	\item The next question to be asked is selected.
	\item This question is asked and an answer is obtained.
	\item This answer is inserted into the model.
	\item The model (which provides estimates of the student's skills) is updated.
	\item (optional) Answers to all questions are estimated given the current estimates of student's skills.
\end{itemize}
In this section we will take a closer look on the model structures for different approaches. Also we will discuss the question selection process from step 1 of the list above. Insertion to the model (step 3) and consequent update of the model (steps 4 and 5) is always done with respective tools for the particular model and will not be extensively discussed here. This topic, especially for BNs, is covered in~\cite{Plajner2015}. We performed experiments on empirical data with different models of following model types. Results of these experiments are not a part of this paper but they are available in~\cite{Plajner2016}.

\section{Building Models with the Help of IRT}
\label{sec_IRT}
The beginning of Item Response Theory (IRT) stems back to about 5 decades ago~\cite{Lord1968, Rasch1960, Rasch1993}. This approach is different from the Classical Test Theory (CTT) and it is getting scientific attention ever since.  IRT allows more specific measurement of certain abilities of an examinee. Internationally, there is a large amount of tests adapting this concept. It has stronger assumptions but it also provide stronger results. Nevertheless, its spread is not as high as could have been expected. This smaller impact might be caused by the fact that there is a requirement for a stronger statistical and theoretical preparation of a test creator than in the CTT. In the Czech Republic there is just a few large normalized tests which use this concept~\cite{2011psychometrics}\footnote{One of them is, for example, the Woodcock–Johnson test~\cite{Anton2010}}. 

\begin{figure}
\centering
  \includegraphics[width=0.8\columnwidth]{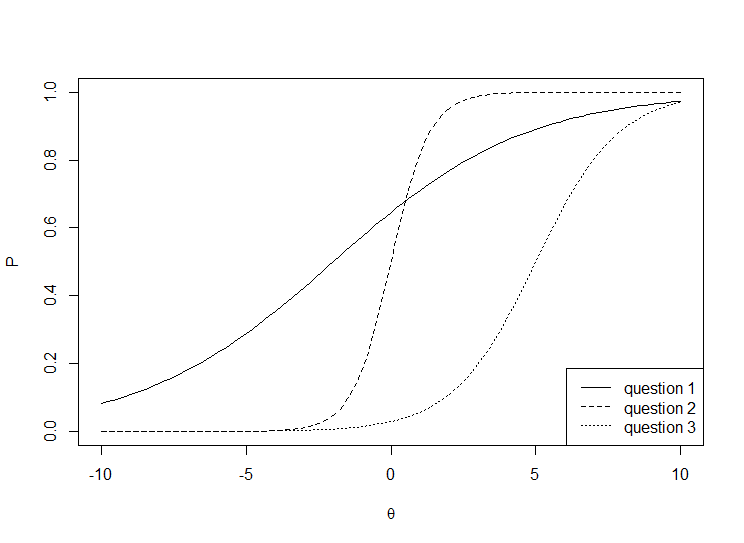}%
  \caption{Item Response Functions}%
	\label{pic:IRFs}%
\end{figure}

IRT expects a student to have an ability (skill) which directly influences his/her chance of answering a question correctly. This ability is called latent ability or latent trait $\theta$. When we have only one variable\footnote{There are variants of multidimensional IRT model where it is possible to have more then one latent variable but in this section we are going to discuss only models with one latent variable.}, it is common to refer to it as proficiency variable. We will stay with the more general skill variable term because we will have more variables in the following parts (Bayesian networks). Every question of the IRT model has an associated item response function (IRF) which is a probability of a successful answer given $\theta$. There are more variants of the shape of this IRF but mostly a 3 parametric model is used (often called 3PL). These parameters reshape a standard logistic function. The resulting IRF, as the probability of a correct answer to \textit{i-th} with the ability of $\theta$, is given by a formula
$$
p_i(\theta) = c_i + \frac{1-c_i}{1+e^{-a_i(\theta-b_i)}}
$$
where $c_i$ is a parameter for guessing, $a_i$ sets the scale of the question (this sets its discrimination ability - a steeper curve better differentiate between students), $b_i$ is the difficulty of the question (horizontal position of a curve in space). An example of typical IRFs is shown in the Figure~\ref{pic:IRFs}. The dependence of a correct answer probability $P$ based on the skill $\theta$ is displayed. The Table~\ref{tab:IRFs} shows parameters of these functions.

IRFs are created either by setting parameters manually or automatically through machine learning procedures. Manual creation is done by field experts based on their knowledge and experiences in the field. Automatic creation is done from collected data as most likelihood estimates of IRFs' parameters.

\begin{table}%
\centering
	\begin{tabular}{lccc} \hline
		Question & a & b & c \\ \hline
		1 & -2 & 0.3 & 0\\
		2 & 0 & 1.5 & 0\\
		3 & 5 & 0.7 & 0\\
	\hline
  \end{tabular}
  \caption{IRFs' parameters}%
	\label{tab:IRFs}%
\end{table}

\subsection{Adaptive Test Procedure}
\begin{figure}%
\centering
\includegraphics[width=0.8\columnwidth]{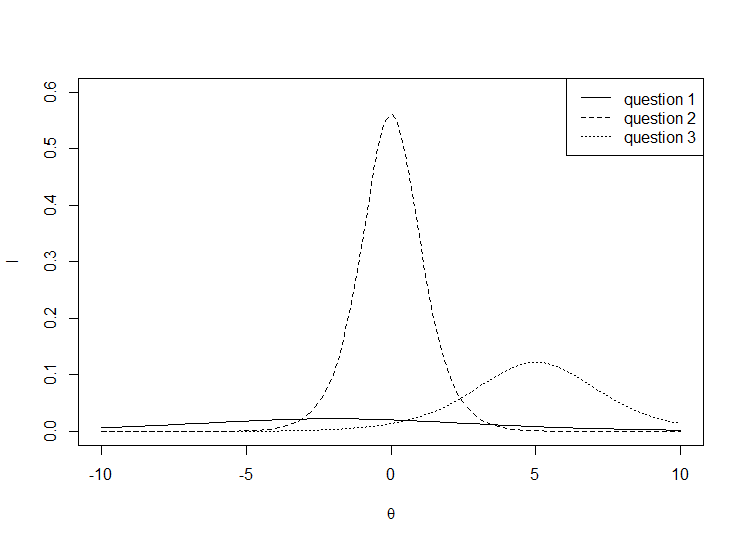}%
\caption{Item Information Functions}%
\label{pic:IG}%
\end{figure}

Building CAT model with IRT is very straightforward. IRT itself, as was described above, is in the form prepared to be used for CAT. With the model fitted from sample data or created by domain experts we have IRFs for every question. In every phase of the test we can compute an estimate of the latent skill $\theta$ based on answers $x$: $p(\theta|x)$. For this estimations Empirical Bayes or Multiple Imputation methods of IRT are used. Knowing the value of the latent skill we know probabilities of a correct answers $p_i(\theta)$ and an incorrect answers $q_i(\theta)$ to every question\footnote{With 3 parametric model these two numbers do not necessarily sum to 1}. More importantly, we are able to calculate the information provided by asking the question. This is called item information and it is given by the formula
$$I_i(\theta)=\frac{(p_i'(\theta))^2}{p_i(\theta)q_i(\theta)}$$
where $p_i'$ is the derivation of the item response function $p_i$. There is an example of typical item information functions (with the same parameters of items as in the Table~\ref{tab:IRFs}) in the Figure~\ref{pic:IG}. This item information provides one, and most straightforward, way of the next question selection. In every step the question $X^*$ which is selected is one with the highest item information. 
$$X^*(\theta) = \arg\max_i I_i(\theta)$$
This approach minimizes the standard error of the test procedure~\cite{Hambleton1991} because the standard error of measurement $SE_i$ produced by $i-th$ item is defined as
$$SE_i(\theta) = \frac{1}{\sqrt{I_i(\theta)}}.$$
This means that the better precision of difficulty we are able to achieve while asking questions the less error of measurement.

\section{Building Models with the Help of BN}

In this section we go over the basic definitions of Bayesian networks, more details can be found in~\cite{Jensen2007, Kjrulff2008}. The use of BNs in educational assessment is discussed in~\cite{Almond2015, Culbertson2015, Millan2010}. This section is focused on the creation Bayesian networks models for CAT. This topic is also discussed, for example, in~\cite{Vomlel2004, Vomlel2004a}.

Bayesian network is a probabilistic graphical model, a structure representing conditional independence statements. It consists of the following: 
\begin{itemize}
	\item a set of variables (nodes),
	\item a set of edges,
	\item a set of conditional probabilities.
\end{itemize}
Edges between variables have to form a directed acyclic graph (DAG). Each variable has a list of mutually exclusive states. For each variable a conditional probability distribution conditioned by its parents is defined, e.g., variable $A$ with parents $B_1$,$B_2$,...,$B_n$ has the conditional probability table\footnote{Note that the variables with no parents have the table in the form $P(A)$} ${P(A|B_1,B_2,...,B_n)}$. 

To build a BN model for adaptive testing we need to perform 3 steps:
\begin{enumerate}
	\item define nodes of the BN,
	\item define connections between nodes, and
	\item specify initial values of conditional probability tables.
\end{enumerate}
 
\subsubsection{Types of Nodes}
We will divide nodes of a BN into three sets. 
\begin{itemize} 
\item A set of $n$ variables we want to estimate $\{S_1,\ldots,S_n\}$. These variables represent latent skills (abilities, knowledge) of a student.
We will call them skills or skill variables. We will use symbol $\boldsymbol{S}$ to denote the multivariable $\boldsymbol{S} = (S_1,\ldots,S_n)$ taking states $\boldsymbol{s} = (s_{1,i_1},\ldots,s_{n,i_n})$. 
\item A set of $m$ variables representing eventual additional information about the student $\{I_1,\ldots,I_m\}$.  
We will use the symbol $\boldsymbol{I}$ to denote the multivariable $\boldsymbol{I} = (I_1,\ldots,I_m)$ taking states $\boldsymbol{i} = (i_1,\ldots,i_m)$.
\item A set of $p$ questions (math problems) $\{X_1,\ldots,X_p\}$.  
We will use the symbol $\boldsymbol{X}$ to denote the multivariable $\boldsymbol{X} = (X_1,\ldots,X_p)$ taking states $\boldsymbol{x} = (x_1,\ldots,x_p)$.
\end{itemize}

\subsubsection{Skills}
Skill nodes model the student abilities and, generally, they are not directly observable. It means they are hidden variables of the model and their value is not known prior to the model creation. Several decisions concerning skill nodes are to be made during the model creation.
 
The first decision is the number of skill nodes itself. Should we expect one common skill or should it rather be several different skills each related to a subset of questions only? In the later case it is necessary to specify which skills are involved in solution of each particular question (i.e. a math problem). These skills become parents of the considered question. Possible relations between them are discussed in the last chapter.

This way we create variables with a given meaning (specific student ability). It is not possible to cover all the necessary skills to solve a question. Also, there are some other aspect important for a question's solution. During the interpretation of a CAT result we have to be careful. Even though we have given the variable the meaning, it is possible that the model learned a combination of this meaning with other factors. Nevertheless, if the model was properly constructed the meaning of the variable should converge to the intended meaning. For example, we have two skills overlapping over some questions and not overlapping over other questions. It means that a skill 1 is needed to solve some questions, skill 2 is needed as well to solve a subset of them and then some others. If we select a student who has (by the model) only one of these skills, we should be able to observe that he/she answered correctly only to a part of questions of the corresponding skills. If this is true and we can specify the skill needed to solve this type of questions, then it is reasonable to assume that the skill interpretation converges to the intended one. 

Another decision we are facing is about the size of a state space of skill nodes. As an unobserved variable, it is hard to decide how many states it should have. Another alternative is to use a continuous skill variable instead of a discrete one but we did not elaborate more 
on this option. For BNs no suitable apparatus to handle continuous parents exists. It would be possible though to create different kind of models with continuous parents\ignore{\footnote{We will elaborate more on this in the last chapter.}}. The use of a discrete state space can be in a way viewed as sampling (or discretizing) the continuous skill variable of the student. It may seem reasonable to create many states for each skill variable but each state increases the total number of parameters of the model (the exact rate depends on the structure). This means that these models are too complex. It may be hard to learn a statistically reliable and stable model if the complexity is high. Conditional probability tables may end up very sparse and that limits the generalization ability of the BN. 

Skills are ordinal variables.A variable $S_i$ with possible states $\{s_{i,1},\ldots,s_{i,n_i}\}$ and an arbitrary state $s_{i,0}$ which is one but first state of the variable, has a probability
$$P(S\leq s_{i,j})$$
of the variable $S_i$ being in one of states $s_{i,k}, k\in \{1,\ldots j\}$ (previous to $j$).  We define
$$P(S_i\leq s_{i,0}) \defeq 0$$ $$ P(S\leq s_{i,n_i}) \defeq 1.$$ 
Then for the probability of the ordinal variable $S_i$ being in the state $s_{i,j}$ the following has to be satisfied:
$$P(S_i = s_{i,j}) = P(S_i \leq s_{i,j}) - P(S_i \leq s_{i,j-1}).$$

If this assumption would not be taken into account it may cause inconsistent results. For example, consider a student answering to a question that is dependent on one skill (with 3 states). Without the ordinality assumption a BN model could result in a following probability distribution: 
\begin{itemize}
	\item With a low level of the skill the chance of a correct answer is high.
	\item With a medium level of the skill the chance of a correct answer is low.
	\item With a high level of the skill the chance of a correct answer is high.
\end{itemize}
This situation is impossible with our definition of a skill to be a reasonable requirement for the question's solution. If the probability of a correct answer is high with the low level of the skill it has to be higher for the medium level and even higher for the high level\footnote{It would be possible to have an inverse situation if we had a skill which negates a correct solution, but it still would have to be monotonic.}. 

\label{observed_score}
It is also possible to use a different BN model where unobserved skill variables are replaced by observed variables. The easiest way is to introduce the total test score as a variable into the model. To do this it is necessary to use a coarse discretization. At first, total scores are divided into $n$ groups and by that we obtain an observed variable having $n$ possible states. The states represent a group of students with similar scores achieved. During the learning phase the variable is observed and the information is used for learning. On the other hand, during the testing, the resulting score is not known -- we are trying to estimate the group into which a test subject falls. In the testing phase the variable is again hidden (unobservable).

Combinations of both types of skill variables are also possible.

\subsubsection{Information about a Student}
Information nodes gather additional information we have about a student. They are observed variables. The number of their states correspond to the possible options of the specific piece of information. One state, labeled as ``unknown'', may be also included for missing values. This should be done especially if it has any information value in the context of the test. For example, if a student does not enter his/her grade of mathematics it may indicate that he does not feel very confident. If it is the case, then this information can be used to our advantage. 
This additional information may improve the quality of the student model. 

%As we discuss in the following chapter, its added value showed up as not very essential. The inclusion of these variables makes the model more complex (more parameters need to be estimated). It may mislead the reasoning by creating prejudices about certain groups of students. The added benefit is low especially in the later stages of testing when sufficient information about a student is collected from his/her answers. 

\subsubsection{Questions}

The last type of nodes are question nodes. This node type holds answers to individual questions. Its state space depends on the number of possible answers to a question. As it was already mentioned it is difficult to build a computerized system for evaluation of answers which do not use the multiple choice question type. In some cases it may be possible to have open answers to questions but in most cases these would be too hard to process. With multiple choice, a question node has two possible state spaces:
\begin{enumerate}
	\item one state for each possible answer,
	\item one state for the correct answer and one for any incorrect answer.
\end{enumerate}
The former case is more informative. It gives us a possibility to differentiate between students not only based on the fact that the answer is correct/incorrect but as well on the fact which incorrect one it is. Nevertheless, it has some limitations. The more the states the higher the number of model parameters to be learned. With a limited training data it may be difficult to reliably estimate model parameters. It requires larger data set to learn from. 

Another aspect is the concept of fairness. It is questionable if it is fair to make distinctions based on wrong answers. On one hand, a classical test usually do not do this. If the answer is wrong then it does not matter which one it is. On the other hand the selected answer brings additional information about the student's ability and there is no theoretical obstacle why not to use it.

\subsubsection{Connections between Nodes}

The last step in the BN model creation is to define a set of arcs between variables (nodes), i.e., network structure. This set defines relations between skills, questions, and additional information, eventually, also inbetween them. This task is usually done by the domain expert by hand. There are algorithms for automated structure learning, but these algorithms are for general cases and usually do not provide usable results for this specific purpose. We discuss the automated creation more in the last chapter. The expert who is creating a structure has to pinpoint which variables should be connected. Usually, we connect skill nodes with questions in a way that skill nodes are parents of questions when the skill is needed to a question's solution. The connection can be of
\begin{itemize}
	\item a deterministic relation, where we have to assign all the values of an associated conditional probability table (discussed in the next section), or
	\item a specific relation, e.g., AND, OR, etc. (discussed in the last section).
\end{itemize}   

Then, there may be connections between skills if we want to further specify them. For example, we can create a common skill that is brought down to two sub-skills. This common skill have a connection to the two sub-skills, but possibly no questions. If we want to include some additional information the expert also has to define what it influences. It depends on the type of the piece of information. It can influence a skill, or even a question directly if it is an important factor for its solution.

\subsection{Model Learning}
The last action to complete a BN is to define conditional probability tables (CPT) for each node. Values in CPTs represent probabilities of the variable being in a state conditioned by a configuration of its parents for every state of the variable and every combination of its parents. First, we manually input values into CPTs. Values should reflect a general expectation and are created with expert knowledge in the field of the test. These probabilities serve as a starting point for the learning algorithm. Next we learn the model with the standard EM algorithm using collected data. This operation modifies values in CPTs to better reflect the data.

BNs have a large advantage since they can learn from missing data (with some unknown values). The EM algorithm, that is used for learning, has no problems operating with missing data. Also, during the prediction process unknown values are simply not inserted into the network and the prediction is performed without this knowledge.

\subsection{Adaptive Test Procedure}
During the adaptive test we use standard BN inference methods to update the network. These methods estimate probabilities of skill variables as well as probabilities of a success in unanswered questions.\\
One task to solve during the CAT procedure is the selection of the next question. It is repeated in every step of the testing and it is described below.

Let the test be in the state after $s-1$ steps. It means that $s$ questions were already answered and they form the evidence $e$:
\begin{eqnarray*}
e = \{X_{i_1} = x_{i_1},\ldots,X_{i_n} = x_{i_n} | i_1,\ldots,i_n \in \{1,\ldots,m\}\}. 
\end{eqnarray*}
Remaining questions
\begin{eqnarray*}
\mathcal{X}_s = \boldsymbol{X} \setminus e
\end{eqnarray*}
are unobserved (unanswered).

The goal is to select a question from $\mathcal{X}_s$ to be asked next. 
We select a question with the largest expected information gain. 

We compute the cumulative Shannon entropy over all skill variables of $S$ given evidence $e$.
It is given by the following formula:
$$
H(e) = \sum_{i = 1}^n \sum_{j = 1}^{i_n} -P(S_i=s_{i,j}|e) \cdot \log P(S_i=s_{i,j}|e).
$$

Assume we decide to ask a question $X' \in \mathcal{X}_s$ with possible outcomes $x'_1,\ldots,x'_p$. 
After inserting the observed outcome the entropy over all skills changes. 
We can compute the value of new entropy for evidence extended by $X' = x_j'$, $j \in \{1,\ldots,p\}$ as:
\begin{eqnarray*}
H(e,X'=x_j') = \sum_{i =1}^n \sum_{j=1}^{n_i} \begin{array}{ll}
-P(S_i=s_{i,j}|e,X'=x_j')\\
\cdot \log P(S_i=s_{i,j}|e,X'=x_j') 
\end{array}\enspace .
\end{eqnarray*}
This entropy $H(e,X'=x_j')$ is the sum of individual entropies over all skill nodes. Another option would be to compute the entropy of the joint probability distribution of all skill nodes. This would take into account correlations between these nodes. In our task we want to estimate marginal probabilities of all skill nodes. In the case of high correlations between two (or more) skills the second criterion would assign them a lower significance in the model. This is the behavior we wanted to avoid. The first criterion assigns the same significance to all skill nodes which is a better solution. For our problem, the greedy strategy based on the sum of entropies provides good results. Moreover, the computational time required for the proposed method is lower.

Now, we can compute the expected entropy after answering question $X'$: 
\begin{eqnarray*}
EH(X',e) & = & \sum_{j=1}^p P(X'=x_j'|e) \cdot H(e,X'=x_j') \enspace .
\end{eqnarray*}
Finally, we choose a question $X^*$ that maximizes the information gain $IG(X',e)$
\begin{eqnarray*}
X^* & = & \operatorname*{arg\,max}_{X' \in \mathcal{X}_s} IG(X',e) \ , \ \mbox{where}\\
IG(X',e) & = & H(e)  - EH(X',e) \enspace .
\end{eqnarray*}

\subsection{Obtaining Total Score from Skills}
\label{sec:converting_skill}
BN models usually produce estimates of student skills. In some cases this is more useful than a regular score. On the other hand if we want to obtain a score in terms of achieved points we have to transform these skills. First, we define a score $SC$ as a weighted sum of skills ($S_1 = s_1,\ldots,S_n = s_n$):
$$SC \defeq \sum_{i=1}^n s_i C_i$$
$C_i$ is a weight associated with the $i$-th skill. These weights define the maximum score
$$SC_{max} = \sum_{i=1}^n s_{i,n_i} C_i,$$
where $s_{i,n_i}$ is the last possible state of $S_i$.

The weights $C_i$ can be set to any value. There are two special cases. The first is, when all the weights are set to be equal
$$C_i = C.$$
Then the impact of each skill on the total score depends on the number of the skill's states. The second is, when we want all the skills to have the same impact on the score. Then weights have to be set to
$$C_i = \frac{n_{max}}{n_i} C,$$
where $n_{max} =  \max\limits_{i}  n_i$ and $C$ is a scaling constant.

During the testing process, the states of skills $S_1,\ldots,S_n$ are unknown. We use their estimates to compute an expected value of the total score:
$$E(SC) = \sum_{i=1}^n\sum_{j=1}^{n_i}P(S_i = s_{i,j})s_{i,j} C_i$$
For a tested student this expected total score is our estimate of the real total score.
\section{Building Models with the Help of NN}
Neural networks are models for approximations of non-linear functions. We present a brief overview of NNs. For more details about NNs, please refer to~\cite{haykin2009, aleksander1995}. 

There are three different parts of a NN:
\begin{enumerate}
	\item an input layer,
	\item several hidden layers, and
	\item an output layer.
\end{enumerate}
Each layer consist of several nodes called neurons. These neurons have connections to the next layer. Usually the connections are formed from one neuron to all neurons in the next layer. Every connection has an associated weight. These weights are used to calculate a value of a neuron from values of its predecessors. A substantial difference between NNs and previously described IRT and BN models is that NNs are learned by supervised learning. In this scenario an output (result), that is to be predicted by a NN, has to be known during learning. IRT and BNs are usually working with unsupervised learning~\cite{schlesinger2002}. For them we do not need to know the output. That allow us to learn models for unknown values of skills. During the learning it is not necessary\footnote{BNs are able to learn even in a supervised fashion knowing the output value} to input other values than question responses. This pattern does not work with neural networks. It is necessary to provide target values during the learning algorithm. In this case we can use score results of the test as target values. It means, that the NN model can not predict the skill of a student but it can predict his/her score directly.

The input layer of the NN model for CAT is created by as many nodes as the number of questions is. For every question we are feeding its result into the neuron. We have two options how to encode information about the answer. It is either 1 (or 1,...,n if it is possible to have more points) for a correct answer and 0 for an incorrect one. Inserting 0 to a node means that there will not be any activation of such node. If we want to activate it even for an incorrect answer we have to encode it as -1. 

There is a general problem of missing data with NNs~\cite{Hastie2009,Pesonen1998}. In order to produce a result NN has to obtain values to all its nodes. There are many different methods to overcome this problem. In our research we input either a value of 0 (there wont be any activation of a neuron then) or an average score for a question (hopefully producing an average result from that question).

The number and the size of hidden layers is up to our choice. There is no specific rule how to choose the best specifications of a NN.

The last output layer contains only a single node. The value of this node corresponds to predicted score of a student.

Learning of the NN model is done by a standard back propagation algorithm from collected data.

\subsection{Selecting the Next Question}
In the two previous models we have used the entropy reduction criterion to select the next question. The entropy was measured on the skill variables. We have no skill variables with NN, only a score output. Measuring an entropy in this case is not possible, because reducing the entropy of total score would mean that we are trying to push a student to some specific score value. With score there is no reason for this, with skills we wanted a student to reach a certain level of skill. Instead, we propose a simple criterion to deal with the selection of the next question. We want the selected question to provide us as much information as possible about the student. That means that a student who answers incorrectly should be as far as possible on the score scale from the one who answers correctly. Let the $SC|X_{i,x}$ be the score prediction after answering the $i$-th question's state $x$, $P(X_{i,x})$ the probability of state $x$ to be the answer to question $i$. $P(X_{i,x})$ can be obtained, for example, by statistical analysis of answers. We select a question $X^*$ maximizing the variance of predicted scores:

$$X^* = \arg\max_i \mathrm{V\underset{x}ar} (SC|X_{i,x}) = \sum_x P(X_{i,x})(SC|X_{i,x} - \overline{SC|X_i}) , $$
where $$\overline{SC|X_i} = \sum_xP(X_{i,x}){SC|X_{i,x}}$$
is the mean values of predicted scores.

\section{Some Remarks on Models}
\subsubsection{Scoring}
Both IRT and BN models usually estimate the skill of a student. We have to perform transformations to the score scale if we want to produce a score of the test. For BN it was discussed in the Section~\ref{sec:converting_skill}. For IRT models a similar procedure can be applied. NN models predict the resulting score directly thus there is no conversion needed.
\subsubsection{Question nodes} BN allow us to exploit every answer to a question as an information about the student. This may help the adaptive test to evolve faster in case there are some answers which are ``more'' wrong that other wrong answers. As was mentioned above it requires large data samples for learning to avoid overfitting.\\
\subsubsection{Additional Information} BN and NN models allow us to include additional information about a student. This is not possible in the standard version of IRT-CAT. 
%The inclusion of such additional information may improve prediction power of the model especially during the first stages of the test, but its influence rapidly decrease in the later stages. The use of this information is also questionable in terms of fairness. It may be hard to justify results of the test being influenced by factors other than answers. On the other hand if we want to build a tutoring system, there should be no problem using it.

\chapter*{Future Work} \addcontentsline{toc}{chapter}{Future Work}
In this section we present a brief overview of research problems of our further interest. 

\subsubsection{Constrained question selection}
Adaptive tests constructed with the IRT model have various well established methods of selecting next questions. As it was discussed in the first chapter, the selection process should be carefully managed to prevent the overuse of some questions as well as very similar question combinations for many participants. One way IRT researchers are solving this issue is through a series of constraints in the selection process~\cite{VanderLinden2004,Stocking2000}. We plan to analyze the use of such constraints for BN models. The research will be conducted in order to define the correct constraints, their application and their impact on the CAT procedure.

\subsubsection{Precision of skills measurement}
Stopping rules for CAT used in IRT are of two kinds. \begin{itemize}
	\item Practical, e.g., limited time of the test or a specific number of asked questions, or
	\item Statistical, e.g., reliability of the test (precision of measured latent skill).
\end{itemize} 
The goal of this research path is to provide a similar criteria for BN-CAT models as well. We will establish a criterion able to provide a statistically sound precision of estimates. Afterwards, we will review the effect of using this criterion as the stopping rule for CAT.

\subsubsection{Model quality criterion}
So far in our research we were evaluating a model quality based on its predictions of answers to remaining questions. This criterion provide reasonable results but it also at the same time has some drawbacks. First of all, for some models similar to each other it is often hard to order them in terms of quality. The prediction accuracy varies over the test run and the order of models changes in different steps (different numbers of asked questions). Next, some models are able to predict the student's skills very well while their prediction power back to answers is not that good. These models have a disadvantage compared to other models (with better backwards precision) if we use the current criterion. This applies especially for some of neural networks models. 

We will focus on the design and testing criteria that are able to take into account more aspects of models for CAT. We would like to better describe, and experimentally verify on data, criteria that work with skills estimates. Specifically, we will measure the quality of a model as a correlation of the predicted score with the real score. Also we want to compare the final ordering of students based on predictions and real values. 

\subsubsection{Model creation}
One reason of the small spread of the CAT use is in the high amount of work required to create a student model. The process itself may be very useful due to its highly organized character and statistically strong results, but it requires a certain level of expertise in modeling and statistics. 

We want to address an option of an automated model creation based on data. Algorithms for BN structure learning in a general case exist, some examples are in~\cite{Margaritis2003}. These algorithms cover general structure learning which may serve well in some areas but in case of CAT it usually does not reflect our situation very well. We would like to explore special types of models which would fit better for our needs as the CAT model. This leads to learning a model which contains a local structure. It is an interesting theoretical task for the whole BN community, not limited only to CAT.

\subsubsection{Local structure in BNs}
\begin{figure}%
\centering
\begin{tikzpicture}
  \SetGraphUnit{2.5}
	\SetUpEdge[style={->,>=biggertip}]
	\Vertex[Math,L=X_2] {X2}
 
  \EA[Math,L=X_3](X2){X3}
  \WE[Math,L=X_1](X2){X1}
	\SO[Math,L=Y](X2){Y}
  \Edge(X2)(Y)
  \Edge(X1)(Y)  
	\Edge(X3)(Y)  
	
\end{tikzpicture}
\caption{Simple Bayesian network}%
\label{fig:net}
\end{figure}
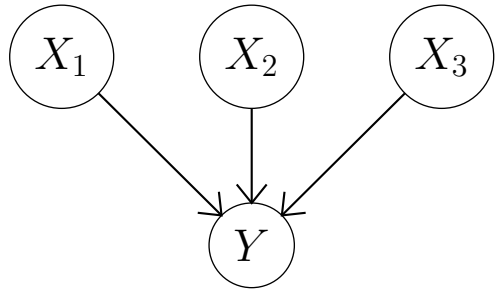

Bayesian networks encode conditional probabilities. These probabilities can be encoded in many ways. One of the most common is to define a conditional probability table (CPT). This table defines a conditional probability of a variable for every combination of its parents. For example, let us consider the network in the Figure~\ref{fig:net}. The CPT for this network with a child $Y$ and parents $X_1,X_2,X_3$ is in the Table~\ref{tab:cpt} (values are chosen at random). In the general case we have to specify $2^n$ parameters where $n$ is the number of parents. In the example case, we need $2^3 = 8$ parameters. This amount of parameters correspond to binary nodes. For nodes with more than two states, the total amount of parameters is even higher.

\begin{table}%
\centering
\begin{tabular}{lcccccccc}
\hline
$X_1$ & 1 & 1 & 1 & 1 & 0 & 0 &0 &0 \\
$X_2$& 1 & 1 & 0 & 0 & 1 & 1 &0 &0 \\
$X_3$& 1 & 0 & 1 & 0 & 1 & 0 &1 &0 \\
\hline
$Y(0)$ & 0.05 & 0.41 & 0.12 & 0.67 & 0.85 & 0.5 &0.9 &0.08 \\
$Y(1)$ & 0.95 & 0.59 & 0.78 & 0.33 & 0.15 & 0.5 &0.1 &0.92 \\
\hline
\end{tabular}
\caption{CPT for the BN in~\ref{fig:net}}
\label{tab:cpt}
\end{table}

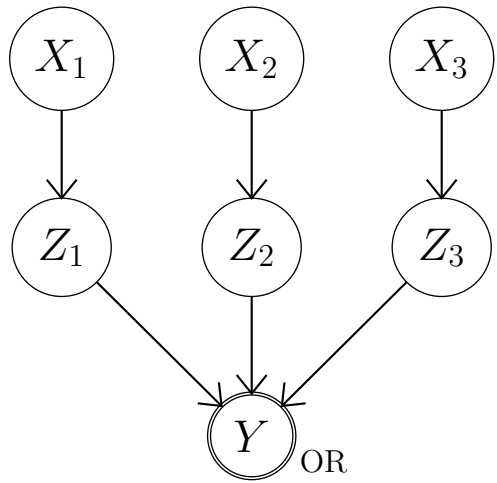
\begin{figure}%
\centering
\begin{tikzpicture}
  \SetGraphUnit{2.5}
	\SetUpEdge[style={->,>=biggertip}]
	\Vertex[Math,L=X_2] {X2}
 
  \EA[Math,L=X_3](X2){X3}
  \WE[Math,L=X_1](X2){X1}
	\SO[Math,L=Z_1](X1){Z1}
	\SO[Math,L=Z_2](X2){Z2}
	\SO[Math,L=Z_3](X3){Z3}
	\SO[Math,L=Y,style = {double}](Z2){Y}
	\node	at ([shift=(-7cm:-1cm)]Y) (l) {OR};	
  \Edge(X2)(Z2)
  \Edge(X1)(Z1)  
	\Edge(X3)(Z3)  
	\Edge(Z2)(Y)
  \Edge(Z1)(Y)  
	\Edge(Z3)(Y)

\end{tikzpicture}
\caption{Bayesian network with noisy OR}%
\label{fig:net2}
\end{figure}

If we have additional information about the structure (relations) of BN variables we can use this information to our benefit. The local structure in a BN allow us to specify these relations and encode conditional probabilities efficiently. The local structure concept is sometimes called as canonical models. A more thorough introduction to the theory of canonical models can be found in~\cite{Diez2007}. Basically, we establish a function that prescribes how to compute a child's value from its parents. The function can be of many different types, but for our illustrative example, we will now consider the noisy OR function only. If there is OR local structure (without noise) it means, that the value of $Y$ is:
$$Y = X_1 \vee X_2 \vee X_3$$
Because the relation is encoded directly in the formula, there is no need to specify a CPT. In this case, we need to know only the values of $X_1,X_2,X_3$. To this model we introduce noise by adding auxiliary variables $Z_1,Z_2,Z_3$. The network then changes from the one in the Figure~\ref{fig:net} to the one in the Figure~\ref{fig:net2}. In this case we need to include probabilities 
$$P(Z_i|X_i)$$
which specify the noise. This forms the noisy OR local structure. The total number of parameters that we have to specify is $n$ (there might be one additional inhibitor variable with connection to each $Z_i$ making it to $2n$ parameters). In our example we have to specify 3 or 6 parameters. It means there is a difference of $O(n)$ for models with a local structure compared to $O(2^n)$ for models without a local structure.

We will explore learning strategies for BNs with a local structure. It means, we have to modify the learning process of the general BN to this special case. During the general structure learning we use criteria for the model ranking. AIC/BIC\footnote{Akaike Iformation Criterion/Bayesian Information Criterion} criteria are popular. These criteria takes into account the prediction quality of a model, but also penalize it for its size. It is necessary to adapt these criteria to the specialized case of BNs with the local structure. AIC/BIC work with the number of nodes to compute parameters, but with the local structure the reduction of number of parameters is essential. 

It is clear that exploiting a local structure in a BN has many advantages. 
\begin{itemize}
	\item First, it is easier to learn a statistically reliable model with less parameters.
	\item We are able to create more complex models where computational operation will be quickly solvable.
	\item It is possible to store this model in less space.
	\item Last but not least, we do not need to specify a large number of conditional probabilities. These probabilities are often obtained from experts and it may be difficult to get reliable estimates of them for large CPTs. With the local structure, we have to specify significantly less conditional probability values.
\end{itemize}

\subsubsection*{Acknowledgements}
The work on this paper has been supported from GACR project n. 16-12010S.

%%%%%%%%%%%%%%%%%%%%%%  Seznam použitých zdrojů  %%%%%%%%%%%%%%%%%%%%%%
\newpage  % SEM NESAHEJTE!

\bibliographystyle{apalike}
\bibliography{studie}

\end{document}